\documentclass[]{aa}

\usepackage{graphicx}
\usepackage{siunitx}
\usepackage{color}
\usepackage{txfonts}
\usepackage{longtable}
\usepackage{mathrsfs}
\usepackage{subcaption}
\usepackage{lscape}
\usepackage{placeins}
\usepackage{subcaption}
\usepackage[colorlinks=true, linkcolor=blue, citecolor=blue, urlcolor=blue]{hyperref}
\usepackage{soul}

\begin{document}

\title{Infrared-to-visible albedo ratio of Phobos}
\subtitle{Comparison with primitive asteroids}

\author{
T.~J.~Dyer\inst{1,2}\thanks{Corresponding author: \email{tdyer@oca.eu}}
\and
J.~Beccarelli\inst{3}
\and
M.~Delbo\inst{1,4}
\and
C.~Avdellidou\inst{4}
\and
M.~Pajola\inst{3}
\and
A.~Milton\inst{5,4}
\and
A.~Lucchetti\inst{3}
\and
G.~Munaretto\inst{3}
}

\institute{Universit\'e C\^ote d'Azur, CNRS--Lagrange, Observatoire de la C\^ote d'Azur, CS 34229 -- F 06304 NICE Cedex 4, France
\and
Centre national d'\'etudes spatiales (CNES), 2 Place Maurice Quentin, 75039 Paris, France
\and
Istituto Nazionale di Astrofisica (INAF) – Osservatorio Astronomico di Padova (OAPd), Vicolo dell’Osservatorio n.5, Padova, Italy
\and
University of Leicester, School of Physics and Astronomy, University Road, LE1 7RH, Leicester, UK
\and
University of Leicester, School of Natural Sciences, University Road, LE1 7RH,
Leicester, UK
}

\date{Received September 30, 20XX}

\abstract
{}
{The ratio between infrared and visible albedo, $p_{\mathrm{IR}}/p_{\mathrm{V}}$, provides a simple proxy for the spectral slope of asteroid surfaces between the visible and near-infrared wavelength ranges. The aim of this work is to assess whether the different spectral units of Phobos occupy the same spectral parameter space as the primitive asteroid populations in order to confirm or not its possible asteroidal origin.}
{For each of the three spectral units of Phobos (Red, Transitional and Blue), we extracted averaged, normalised reflectance spectra by comparing normalised reflectance in the visible (0.50--0.60~\SI{}{\micro\meter}) and near-infrared (3.1-3.8~\SI{}{\micro\meter}) wavelength ranges from the Mars Express OMEGA instrument, to compute a proxy for the infrared-visible albedo ratio ($R_P$).
Then, using the database of visible and infrared albedos compiled in the Minor Planet Physical Properties Catalogue (MP3C), we computed asteroid $p_{\mathrm{IR}}/p_{\mathrm{V}}$ values ($R_A$), to compare to those for Phobos.}
{The $R_P$ for the Red, Transitional and Blue regions vary between 2.90 and 3.12, 2.32 and 2.35, 1.64 and 1.85 respectively. These values span a range occupied by primitive asteroid populations, with the Transitional and Red regions coinciding with the high ratio regime within the $p_{\mathrm{IR}}/p_{\mathrm{V}}$ domain, which is commonly associated with red-sloped primitive asteroids, including D- and T-type objects as well as Jupiter Trojans.}
{}


\maketitle
\nolinenumbers
\section{Introduction}

The infrared reflectance properties of atmosphereless minor bodies provide important constraints on their surface composition, regolith structure, and space weathering state. Measurements from the Wide-field Infrared Survey Explorer (WISE) \citep{wright_2010_the} and its extended NEOWISE mission \citep{mainzer_2011_preliminary} have enabled the determination of infrared geometric albedos, $p_{\mathrm{IR}}$, for large samples of these bodies, in particular, asteroids \citep{mainzer_2011_neowise}. Combined with visible geometric albedos, $p_{\mathrm{V}}$, the latter derived by combining WISE and NEOWISE infrared data with absolute magnitudes obtained in visible light from ground-based surveys, provide a powerful diagnostic of surface behaviour across wavelengths through the ratio $p_{\mathrm{IR}}/p_{\mathrm{V}}$.

The $p_{\mathrm{IR}}/p_{\mathrm{V}}$ ratio can be interpreted as a proxy for the spectral slope trend from visible to near-infrared. While $p_{\mathrm{V}}$ reflects the overall brightness of a surface at a wavelength ($\sim$0.55~\SI{}{\micro\meter}), $p_{\mathrm{IR}}$ samples reflectance in the infrared range, corresponding to the WISE W1 and W2 bands, centred at $\sim$3.4 and 4.6~\SI{}{\micro\meter}, respectively \citep{masiero_2014_mainbelt}. At these wavelengths both reflected sunlight and thermal emission may contribute. 

For minor bodies, variations in $p_{\mathrm{IR}}/p_{\mathrm{V}}$ primarily trace differences in spectral slope between the visible and infrared, and therefore provide information not only of surface composition \citep{demeo_2009_an}, but also on their age as they redden due to space weathering effects \citep{hapke_2001_space, pieters_2016_space, hasegawa_2022_spectral_evolution_dark}.

Primitive main belt and Jupiter trojan asteroid classes, characterised by red spectral slopes, such as D- and T-type asteroids, are expected to exhibit elevated $p_{\mathrm{IR}}/p_{\mathrm{V}}$ values \citep{fornasier_2007_visible, mainzer_2011_neowise}. High ratios have also been reported for outer Solar System small-body populations, verifying the scenario where these asteroids and Jupiter trojans are thought to be captured trans-Neptunian objects during the giant planets orbital instability \citep{morbidelli2005,levison2009,vokrouhlicky2016}. In contrast, less red populations, such as many C-complex asteroids, tend to occupy lower $p_{\mathrm{IR}}/p_{\mathrm{V}}$ values. Despite these trends, the taxonomic dependence and physical interpretation of the $p_{\mathrm{IR}}/p_{\mathrm{V}}$ distribution remain only partially characterised \citep{masiero_2014_mainbelt}, in part because this parameter provides a compressed representation of spectral behaviour compared to full reflectance spectra.

In this work we compare the spectral diversity observed on Phobos, the larger of the Martian moons, at wavelengths similar to those of WISE/NEOWISE W1 band against populations of primitive asteroids. This is because the origin of Phobos remains debated, with competing hypotheses invoking either the capture of a primitive outer Solar System body \citep{ pajola_2013_Phobos, hansen_2018_dynamical,kegerreis_2025_origin} or formation from debris generated by a past giant impact on Mars \citep{craddock2011, rosenblatt_2011_the}. 

Spectroscopic observations have been carried out and analysed with the primary means of distinguishing between these scenarios. However, despite decades of observations spanning between visible and infrared wavelengths, the interpretation of Phobos' surface remains undetermined \citep{rivkin2002,pajola_2012_spectrophotometric,pajola_2013_Phobos,pajola_2018_Phobos,ballouz2019,takir2022,pajola2025,beccarelli2026}. Indeed, its spectrum is generally featureless, and slightly red sloped, lacking strong diagnostic absorption features that would allow for a unique mineralogical identification \citep{pajola_2018_Phobos}. While this behaviour is broadly consistent with primitive carbon-rich asteroids \citep{demeo2015}, similar spectral characteristics could also be reproduced by other effects such as space-weathering, or fine grained silicate materials associated with impact-generated debris \citep{pieters_2000_space, hapke_2001_space}.

As a result, spectroscopy alone has not provided a definitive constraint on Phobos' origin. Instead, some studies emphasise the relative spectral variations \citep{murchie_1991_color} across the surface, rather than absolute compositional identification as a key diagnostic. Visible and near-infrared observations reveal marked spectral differences, commonly described in terms of Blue, Transitional, and Red spectral units \citep{fraeman_2014_spectral, pajola2025, munaretto_2025_phase,beccarelli2026}. \citet{ballouz2019} suggested that certain orbital processes may refresh parts of the surface of Phobos, exposing less-weathered material in blue regions, while leaving other terrains redder and more strongly affected by space weathering.
Specifically, the Red regions exhibit the steepest slopes and lowest V-band reflectance, which is suggestive of more primitive or space-weathered material. Their variability across different surface regions of Phobos suggests that its Red unit represents a primitive and heavily space weathered surface, while the Blue unit is significantly fresher and possibly exogenous \citep{munaretto_2025_phase,beccarelli2026}.

Although \citet{wargnier2025} recently compared the VISNIR spectral properties of Phobos with asteroid populations and Martian materials, the spectral variability of Phobos has not yet been examined against similarly large asteroid samples using infrared-to-visible albedo ratios. For this reason, we compare the distribution of spectral slopes of asteroids with the spectral slope of the Phobos Red, Transitional and Blue units. Specifically, we use high resolution OMEGA hyperspectral data \citep{beccarelli2026b} to compute spectral slopes and link them to spectral slopes derived from the NEOWISE survey in the case of the $p_{\mathrm{IR}}$ values, in addition to other literature sources for the $p_{\mathrm{V}}$ values, which are consolidated by the Minor Planet Physical Properties Catalogue (MP3C)\footnote{\url{mp3c.le.ac.uk}}. By establishing this comparison, we aim to assess whether the spectral regions of Phobos occupy the same parameter space as primitive asteroid populations, and to evaluate the extent to which infrared-to-visible albedo ratios can be used as a diagnostic link between asteroid surfaces and planetary materials.

\section{Data and Methods}

\subsection{Asteroid sample}

Asteroid physical properties were taken from MP3C\footnote{version 3.0.0-beta.22 released 6\textsuperscript{th} March 2026} which compiles data from literature sources, and enables them to be compared. Specifically, we used infrared geometric albedo, of which the sole data provider is NEOWISE, and visible geometric albedo which is provided by extensive literature sources. For the latter, when more than one value exists in the database, we calculated an uncertainty-weighted mean value. We retained only objects with valid measurements of both quantities and then computed $R_A$ as:
\begin{equation}
    R_A = \frac{p_{IR}}{p_{\mathrm{V}}}.
\end{equation}

We then filtered these so that only low albedo asteroids were included ($p_{\mathrm{V}} < 0.12$) to ensure that we do not include asteroids that might have been misclassified, while also being appropriate for comparison with Phobos, whose brightest terrain has a visible geometric albedo of only $p_{\mathrm{V}} \simeq 0.08$ \citep{fornasier_2024_Phobos}. Therefore, including substantially higher albedo asteroid samples would not provide a physically meaningful comparison with the surface of Phobos.

Next, we retrieved the best taxonomic classification for each asteroid, which is given in MP3C following the methodology of \cite{dyer_2026_the}. Primarily for this work we retrieved asteroids that are best classified as C-complex, dark X-complex (also sometimes indicated as P-types, following previous taxonomies), and the T/D type population in the Bus-DeMeo taxonomy \citep{demeo_2009_an}. There are 780, 794, and 464 objects respectively. Then, separately, we selected asteroids that are classified as Z-type per the Mahlke taxonomy \citep{mahlke_2022_asteroid}, yielding 55 objects before the final data-quality filtering. Of these, 32 have finite $R_A$ values and are included in the comparison against Phobos' terrains presented below. Z-types were considered separately because they are expected to occupy the redder part of the infrared-to-visible reflectance distribution and are not defined in the Bus-DeMeo taxonomy. Our comparison with C-complex asteroids is motivated by the fact that they are primitive objects \citep{demeo2015}.

\subsection{Mars Express OMEGA data}
The dataset used in this work is derived from observations acquired by OMEGA (Observatoire pour la Min\'eralogie, l'Eau, les Glaces et l'Activit\'e), the visible and near-infrared imaging spectrometer \citep{Bibring2004} on board Mars Express of the European Space Agency (ESA).

OMEGA provides continuous spectral coverage on Mars, from the visible to the near-infrared ($\sim$0.35-5.1~\SI{}{\micro\meter}), encompassing the wavelength range over which diagnostic spectral slopes and potential absorption features of primitive materials are expected \citep{Bibring2004}. This wavelength range allows $p_{\mathrm{IR}}/p_{\mathrm{V}}$ estimations for different spectral regions of Phobos and compare them with other minor bodies.

\begin{figure*}[t]
    \centering

    \raisebox{0.675cm}{
        \includegraphics[
            height=7.4cm,
            keepaspectratio
        ]{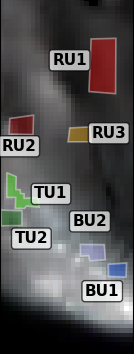}
    }
    \hspace{0.8cm}
    \includegraphics[
        height=8.2cm,
        keepaspectratio
    ]{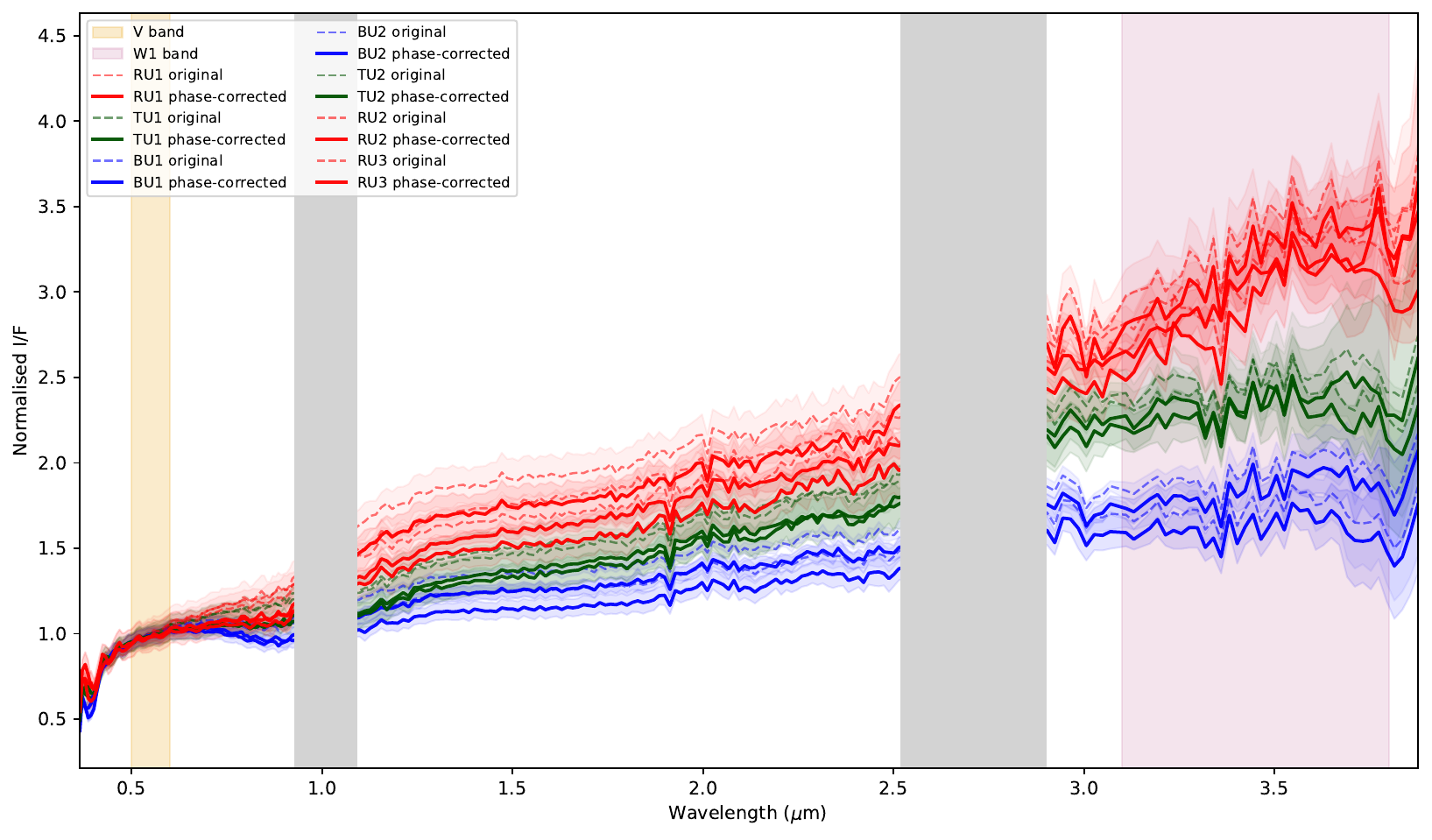}

    \caption{
        Left: Spatial distribution of the ROIs within the Red, Transitional, and Blue units used to extract the data. Right: Original and phase-corrected normalised spectra of the seven ROIs shown on the left. The coloured shaded regions indicate the visible ($0.50$--$0.60~\si{\micro\meter}$) and NEOWISE W1-like ($3.10$--$3.80~\si{\micro\meter}$) wavelength intervals used to calculate $R_{\mathrm{P}}$, while the grey shaded regions mark wavelength intervals excluded because of detector-boundary effects or increased uncertainty. The absorption feature near $3.37~\si{\micro\meter}$ observed in the Blue Unit is discussed by \citet{beccarelli2026b}. All units show a general increase in normalised uncertainty with wavelength, with the Red Unit exhibiting the largest uncertainties, particularly beyond approximately $3~\si{\micro\meter}$.}
    \label{fig:two_images}
\end{figure*}

A key advantage of OMEGA is its ability to obtain spatially resolved spectra over a wide fraction of the surface of Phobos using a single, well-characterised instrument \citep{Bibring2004,gondetbbibringjplangeviny_2010_Phobos}. This enables the construction of a homogeneous spectral dataset, minimising systematic effects arising from instrumental differences or viewing geometry. In contrast to ground-based observations, OMEGA data are unaffected by terrestrial atmospheric absorption and provide broad, continuous spectral coverage. However the interpretation of spectral structure near 1~\SI{}{\micro\meter} remains uncertain: it has been attributed to residual radiometric calibration artefacts in OMEGA data \citep{pajola2025}, although a comparable feature has also been reported in independent datasets \citep{beccarelli2026}. These localised effects are unlikely to significantly affect the broad integrated bands used here because there is no overlap (Fig.~\ref{fig:two_images}).

Furthermore, OMEGA has been used extensively in previous studies of Phobos, allowing direct comparison with earlier analyses of spectral regions and surface heterogeneity. The identification of distinct Blue and Red units, as well as the generally featureless, red-sloped spectra of Phobos, has been robustly established using this instrument \citep{gondetbbibringjplangevinypouletfmurchiesl_2008_Phobos, fraeman2012, fraeman_2014_spectral}. The use of OMEGA data therefore ensures both consistency with the literature and access to one of the most comprehensive spectral datasets currently available for the Martian moons. Further to this point, OMEGA data has a higher spatial resolution than CRISM data.

For Phobos we used spatially resolved spectra acquired by OMEGA, taken on the 22$^\text{nd}$ August 2004 at approximately 151.9~\SI{}{\kilo\meter}, which resulted in a spatial resolution of 171.1~\SI{}{\meter}/px. Using these data we are able to compute broad-band visible-to-infrared reflectance ratios for comparison with $R_A$ values. Because at the Martian heliocentric distance, the thermal excess is dominant with respect to the reflected light for $\lambda > 2.5$~\SI{}{\micro\meter} (see Fig.~4 of \citealt{pajola_2018_Phobos}), this must be estimated and removed. The OMEGA dataset analysed was corrected using the procedure described in \citealp{beccarelli2026b}, which relies on the empirical correction developed by \citealp{clark_2011_thermal}.

For each of the three spectral units (i.e., the Blue, Transitional, and Red units), we selected Regions of Interest (ROIs) representative of each unit following the unit classification of \citet{beccarelli2026b}. The ROIs used in this study are not identical to those analysed by \citet{beccarelli2026b}; instead, they were independently selected (Fig.~\ref{fig:two_images}) to minimise the $\chi^2$ difference between the thermal excess derived from the observed data and the thermal-excess model within each unit, thereby ensuring the most reliable thermal correction.
While the Red unit covers a sufficiently large area to allow the selection of several independent ROIs, the Transitional and Blue units are much less spatially extensive. Consequently, only two representative ROIs were selected for each of these two units.

From each ROI, we extracted the thermal excess-corrected $I/F$ spectra and computed a mean spectrum with associated uncertainties. Results are shown in Fig.~\ref{fig:two_images}, where the uncertainties include the standard error of the mean within each selected ROI, the instrumental noise, and the uncertainty introduced by the thermal correction, as described in Sect. ~\ref{sec:PhobosUncertaintyCalc}. Because $I/F$ is the observed radiance divided by the incident solar flux minus the modelled thermal excess, it therefore retains information on both absolute brightness and spectral shape. The normalised $I/F$ spectra, that we indicate with $\langle I/F \rangle$, were obtained by dividing each mean thermal excess-corrected $I/F$ by its value at the reference wavelength of 0.55~\SI{}{\micro\meter}. This normalisation removes the absolute reflectance level and emphasises the relative spectral behaviour, in particular the visible-to-near-infrared slope. Because our objective is to compare infrared-to-visible spectral behaviour, the band ratios used in this work were derived from the $\langle I/F \rangle$ spectra directly.

We also considered extending this analysis to the \textit{NEOWISE} $W2$ band. In principle, including this longer-wavelength interval would provide additional constraints on the infrared behaviour of the spectral regions. However, this approach was ultimately not retained, as the OMEGA data exhibit significantly increased uncertainties at wavelengths beyond 4.0~\SI{}{\micro\meter}. These large errors limited the reliability of the derived reflectances in this region and prevented a robust comparison with the $W2$ band. Further to this, since the spectra are essentially linear in the near infrared, the difference in band pass should not bring any significant errors.

\subsection{Uncertainty estimation}
\label{sec:PhobosUncertaintyCalc}

The uncertainty estimation for the Phobos spectra consists of three main components: instrumental noise, uncertainty associated with the thermal excess correction, and the statistical uncertainty of the selected Regions of Interest (ROIs).

Since no instrumental uncertainty estimates are provided with the OMEGA dataset, the instrumental noise was estimated from a dark sky region located well away from Phobos in the original observational swath, thereby minimizing contamination from scattered light or target signal. The standard deviation measured in this region was adopted as the instrumental uncertainty at each wavelength.

The uncertainty introduced by the thermal excess correction was estimated using a bootstrap approach. For each pixel spectrum, 100 synthetic realizations were generated by perturbing the measured I/F values according to the estimated instrumental uncertainty, assuming Gaussian-distributed noise. The thermal correction procedure was then independently applied to each realization. The mean corrected spectrum was adopted as the final spectrum, while the standard deviation of the 100 corrected spectra was taken as the uncertainty associated with the thermal correction. This procedure naturally propagates both the instrumental noise and the uncertainty introduced by the subtraction of the fitted thermal emission component.

Finally, the spatial variability within each ROI was quantified through the standard error of the mean spectrum. The total uncertainty shown in the figures was computed by adding in quadrature the bootstrap uncertainty and the ROI standard error,

\begin{equation}
\sigma_{\rm tot}=\sqrt{\sigma_{\rm boot}^2+\sigma_{\rm ROI}^2}.
\end{equation}

At wavelengths shorter than approximately 2.5~\SI{}{\micro\meter}, the total uncertainty is dominated by the instrumental noise and the spatial variability within each ROI. At longer wavelengths, the uncertainty associated with the thermal correction becomes progressively larger, reflecting the increasing contribution of thermal emission to the observed signal.

The spectrum of the Red Unit exhibits a subtle concave curvature beyond approximately 2.5~\SI{}{\micro\meter}. At present, we cannot determine whether the observed curvature is intrinsic to the surface reflectance or a residual effect of the thermal excess correction procedure. The thermal excess has indeed a concave curvature in this wavelength range \citep{david2024}. Also, the uncertainties increase toward longer wavelengths, particularly beyond approximately 3~\SI{}{\micro\meter}, where residual thermal emission effects cannot be excluded. Consequently, the curvature observed beyond 2.5~\SI{}{\micro\meter} should be interpreted with caution.

\subsection{Phase Reddening Correction}
The OMEGA data for Phobos were acquired with a high phase angle of $64^\circ$, while WISE/NEOWISE  observations were acquired at significantly lower phase angles ($15^\circ - 35^\circ$) \citep{masiero2011}. A phase-reddening correction was applied to the normalised OMEGA spectra using the aforementioned phase angle of $64^\circ$. At such viewing geometries, the spectral slope can be affected by phase reddening, where surfaces appear progressively redder as phase angle increases. Since the aim of this work is to compare the intrinsic spectral behaviour of the Phobos units with asteroid populations, this observational geometry effect needed to be estimated and removed before interpreting the infrared-to-visible spectral ratios.

The correction was applied using CaSSIS-derived phase-reddening coefficients \citep{munaretto_2025_phase}. The Mature Red Unit coefficients were combined using an inverse-variance weighted mean, while the Mature Blue Unit coefficient was used directly. The Transitional Unit correction was defined as the average of the Red and Blue corrections, with uncertainties propagated in quadrature. The coefficients were multiplied by the OMEGA observation phase angle of $64^\circ$ to obtain the total correction.

The correction was applied as a linear spectral tilt between the CaSSIS BLU and NIR wavelengths, 0.4999 and 0.9367~\SI{}{\micro\meter}, respectively. The correction was zero at the BLU wavelength, increased linearly to its full value at the NIR wavelength, and was held constant at longer wavelengths. Corrected uncertainties were calculated by adding the original spectral uncertainties and the uncertainty on the phase-reddening correction in quadrature.

The resulting correction is relatively small, consistent with the weak phase-reddening effect measured on Phobos by \citet{munaretto_2025_phase} and with the generally limited phase reddening observed for D-type asteroids \citep{perna2018}.

\subsection{Synthetic V- and W1-band photometry}

Following the phase-reddening correction, we computed synthetic V- and W1-band reflectances for each Phobos spectral unit from the thermal excess-corrected and normalized OMEGA spectra. Rather than calculating a simple arithmetic mean over rectangular wavelength intervals, we computed the filter-weighted mean reflectance by multiplying each spectrum by the Bessell V and WISE W1 relative system response curves and normalizing by the integrated filter response. This procedure yields synthetic band reflectances that are directly comparable to broadband photometric observations.

For each band, the response-weighted reflectance was calculated as

\begin{equation}
\langle R \rangle_{\mathrm{band}} =
\frac{
\int R(\lambda)\,S_{\odot}(\lambda)\,
T_{\mathrm{band}}(\lambda)\,\lambda\,
\mathrm{d}\lambda
}{
\int S_{\odot}(\lambda)\,
T_{\mathrm{band}}(\lambda)\,\lambda\,
\mathrm{d}\lambda
},
\end{equation}

where $R(\lambda)$ is the phase-corrected, normalised OMEGA $I/F$ spectrum, $S_{\odot}(\lambda)$ is the adopted solar spectral shape, and $T_{\mathrm{band}}(\lambda)$ is the relative system response of the corresponding filter. The additional factor of $\lambda$ accounts for the photon-counting response of the detector.

The Bessell V and WISE W1 response curves were obtained using the \texttt{speclite} filter library. The solar spectral shape was approximated using a 5778~K Planck function; its absolute normalisation is unimportant because it cancels in the ratio. The response-weighted effective wavelengths were $0.5526~\mu\mathrm{m}$ for V and $3.3643~\mu\mathrm{m}$ for W1.

For each ROI within each unit, we then defined the infrared-to-visible reflectance ratio as

\begin{equation}
R_P =
\frac{
\langle R \rangle_{\mathrm{W1}}
}{
\langle R \rangle_V
}.
\label{eq:R_P}
\end{equation}

The uncertainties on the synthetic band reflectances were propagated from the wavelength-dependent uncertainties of the OMEGA spectra using the normalised filter-integration weights. The uncertainty on $R_P$ was subsequently obtained using standard propagation for a ratio.

We note that $R_P$ is a ratio of reflectances, but it can be used as a direct proxy of the albedo ratio $R_A$. This is because $p_{IR}$ is assumed by \citealp{mainzer_2011_neowise} to obey the same relationships of $p_V$ apart from a multiplicative factor, which, by construction, is equal to $R_A$. Therefore from here on we call $R_P$ an albedo ratio, despite it being a reflectance ratio.

\section{Results and Discussion}

\subsection{Phobos reference ratios}

The reflectance ratios derived from the OMEGA spectra of Phobos yield the values reported in Table \ref{tab:individual_roi_rp}.

\begin{table}
\caption{Infrared-to-visible reflectance ratios of Phobos units.}
\label{tab:individual_roi_rp}
\centering
\begin{tabular}{l l c}
\hline\hline
ROI & Unit & $R_{\mathrm{P}}$ \\
\hline
RU1 & Red          & $2.90 \pm 0.05$ \\
RU2 & Red          & $3.00 \pm 0.03$ \\
RU3 & Red          & $3.12 \pm 0.07$ \\
TU1 & Transitional & $2.35 \pm 0.05$ \\
TU2 & Transitional & $2.32 \pm 0.04$ \\
BU1 & Blue         & $1.85 \pm 0.02$ \\
BU2 & Blue         & $1.64 \pm 0.03$ \\
\hline
\end{tabular}
\tablefoot{$R_{\mathrm{P}}$ denotes the ratio of the synthetic W1-band
(3.4~\si{\micro\meter}) to V-band (0.55~\si{\micro\meter})
reflectances derived from the phase-reddening-corrected OMEGA spectra.
Uncertainties are $1\sigma$. The Red, Transitional, and Blue units show
progressively decreasing $R_{\mathrm{P}}$ values.}
\end{table}

These values define a clear progression in infrared-to-visible albedo behaviour, from blue material to redder material (in the Transitional Units), to even redder material (in the Red Units). This trend is consistent with the known spectral slope variations across the surface of Phobos \citep{fraeman_2014_spectral, hasegawa_2022_scheila,munaretto_2025_phase, pajola2025, beccarelli2026,beccarelli2026b}. Because the adopted infrared interval overlaps the 3.37~\SI{}{\micro\meter} region analysed in detail by \citealp{beccarelli2026b}, this progression should be interpreted as an integrated broad-band spectral behaviour, including both continuum slope and any spectral structure within the selected wavelength range.

\subsection{Distribution of asteroid \texorpdfstring{$R_A$}{R\_A} values}

\begin{figure*}[t]
    \centering

    \begin{subfigure}{0.49\textwidth}
        \centering
        \includegraphics[width=\linewidth]{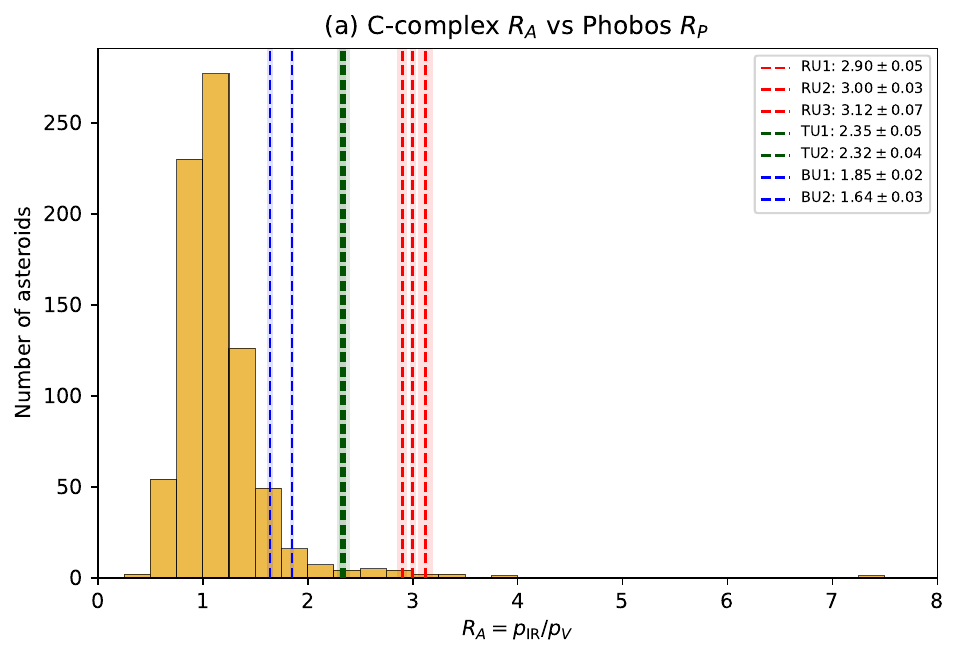}
        \phantomcaption
        \label{fig:a}
    \end{subfigure}
    \hfill
    \begin{subfigure}{0.49\textwidth}
        \centering
        \includegraphics[width=\linewidth]{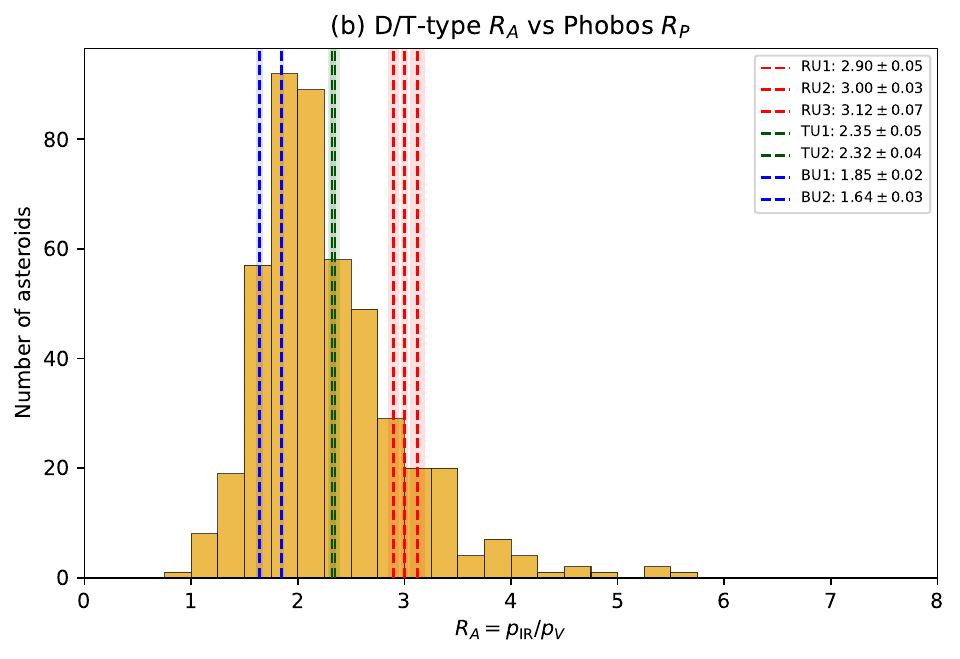}
        \phantomcaption
        \label{fig:b}
    \end{subfigure}

    \begin{subfigure}{0.49\textwidth}
        \centering
        \includegraphics[width=\linewidth]{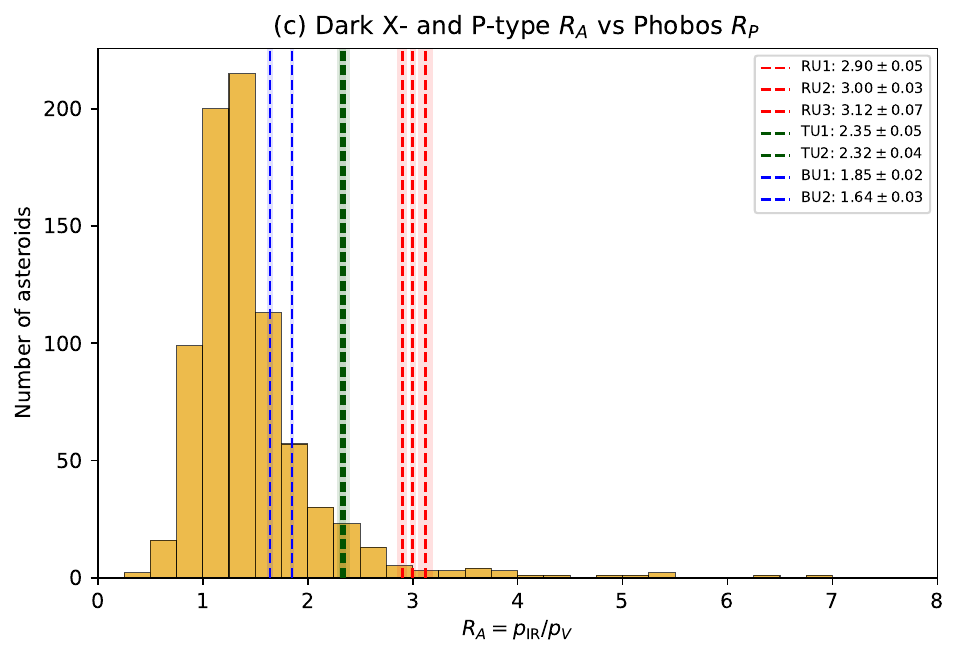}
        \phantomcaption
        \label{fig:c}
    \end{subfigure}
    \hfill
    \begin{subfigure}{0.49\textwidth}
        \centering
        \includegraphics[width=\linewidth]{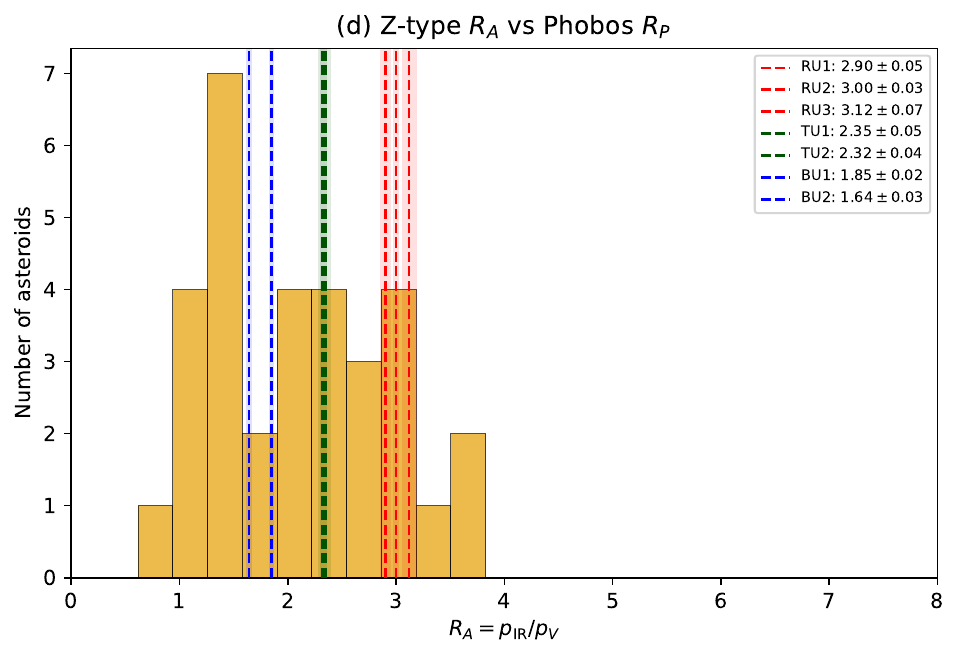}
        \phantomcaption
        \label{fig:d}
    \end{subfigure}

    \caption{Comparison of $p_{\mathrm{IR}}/p_{\mathrm{V}}$ distributions with W1/V reference lines for different asteroid classes.}
    \label{fig:all_populations}
\end{figure*}

The distribution of asteroid infrared-to-visible albedo ratios spans a broad range (Fig.~\ref{fig:all_populations}), with the majority of objects clustering at moderate values and a smaller subset extending toward significantly higher ratios. Primitive taxonomic classes \citep{demeo_2022_connecting}, in particular P/X, T, and D asteroids, preferentially populate the high-ratio tail of the distribution. This behaviour is consistent with their characteristically red spectral slopes extending from the visible into the infrared \citep{demeo_2009_an,fornasier_2011_spectroscopic, elbezsebastien_2026_primitive}.    

In contrast, populations such as the C-complex tend to occupy lower $R_A$ values and exhibit a more compact distribution, indicating comparatively weaker infrared spectral slopes \citep{demeo_2009_an, elbezsebastien_2026_primitive}.

Quantitatively, the mean infrared-to-visible albedo ratios are $\overline{R_A}=1.16\pm0.02$ for the C-complex, $1.47\pm0.02$ for dark X/P-types, $2.12\pm0.15$ for Z-types, and $2.29\pm0.03$ for D/T-types, where the quoted uncertainties are standard errors on the mean. The D/T-type population exhibits the highest mean broad-band ratio, but it is similar (within $1\sigma$) to the Z-types. However, the Z-type sample contains only 32 objects and its $R_A$ distribution is broad, irregular, and comparatively weakly peaked. 

\subsection{Comparison between asteroids and Phobos}
When the $R_A$ values of our asteroid populations are compared to the $R_P$ values of Phobos spectral units, the latter occupy distinct regions within the $p_{\mathrm{IR}}/p_{\mathrm{V}}$ space.

Figure~\ref{fig:a} shows that the C-complex population is concentrated predominantly at $R_A$ values below the Phobos measurements. Its high-$R_A$ tail overlaps the two Blue Unit ROIs, particularly BU2, whereas the Transitional and Red Unit ROIs lie beyond the main C-complex distribution. The dark X/P-type population extends to higher ratios and shows an overlap with both Blue Unit ROIs. The Transitional Unit ROIs intersect the comparatively sparse high-$R_A$ tail of this population, while the Red Unit ROIs lie beyond its upper end.

All the Blue Unit, Transitional Unit and Red Unit ROIs fall within the main D/T-type distribution, although the Blue Unit measurements occur towards its lower-ratio side.

The Z-type sample spans the $R_A$ values measured for all three Phobos spectral units (Fig.~\ref{fig:d}), indicating that these values are compatible with the overall range of the population. However, the overlap is not uniform. Most Z-types are concentrated at lower-to-moderate $R_A$ values, whereas the Transitional and Red Unit reference values fall within the comparatively sparse, nearly flat high-$R_A$ tail of the distribution. Thus, although Phobos-like $R_A$ values occur among Z-types, they are not characteristic of the population. Z-types should therefore not be regarded as the closest or most representative taxonomic analogue on the basis of $R_A$ alone, particularly given the limited sample of only 32 objects.

An additional complication is that Z-types may exhibit concave near-infrared spectra \citep{mahlke_2022_asteroid}. Consequently, a relatively low $R_A$ value does not necessarily correspond to a uniformly shallow spectral slope. Instead, a spectrum may rise steeply at shorter near-infrared wavelengths before flattening or curving downward toward the W1 wavelength range, thereby reducing the broadband $p_{\mathrm{IR}}/p_V$ ratio. The $R_A$ parameter should therefore be interpreted as an integrated visible-to-infrared slope proxy rather than as a direct measure of either the continuum slope or the spectral curvature.

\begin{figure*}
    \centering
    \includegraphics[width=0.9\textwidth]{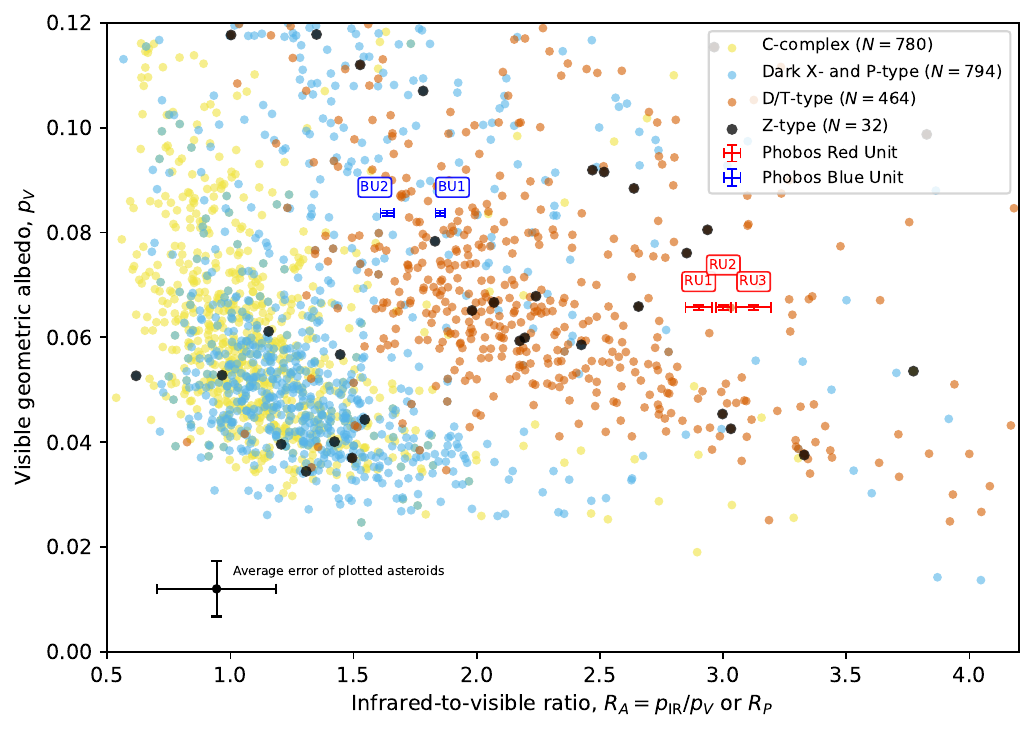}
    \caption{Visible geometric albedo, $p_V$, as a function of the infrared-to-visible albedo ratio, $R_A=p_{\mathrm{IR}}/p_V$, for the selected asteroids, selected on the basis of their spectroscopic classes/complexes compared to the same quantities for the Blue and Red Units of Phobos.}
    \label{fig:albedo_ratio}
\end{figure*}

To complement the comparison based on $R_A$ alone, we examined  geometric visible albedo as a function of $R_A$ for the selected asteroid populations (Fig.~\ref{fig:albedo_ratio}). Estimation of the visible albedo from the OMEGA reflectance is beyond the scope of this paper. Hence, we utilize well established values from \citet{fornasier_2024_Phobos}, their Tab.~4, namely $p_V = 8.37\pm0.05 $\% and  $6.57\pm0.05$\% for the Stickney rim and Stickney floor which correspond to the Blue Unit and the Red Unit. No albedo information is available for the Transitional Unit.

The individual Phobos measurements show where the Blue and Red Unit ROIs overlap with the asteroid distribution in geometric visible albedo as a function of the infrared-to-visible albedo ratio. BU1 and BU2 fall within a region occupied by several primitive asteroid populations, predominantly D/T-types, the sparsely populated edge of the dark X/P-type distribution, and, more marginally, the C-complex. Their relatively high visible albedos and $R_P$ values  between $1.6$ and $1.9$ place them outside the core distribution of the different types of asteroids, while still being mostly compatible with D/T-, Z-types and to some extent to the dark X/P-types.

The three Red Unit ROIs form a compact group at $R_P \simeq 2.9$--$3.1$ and $p_V \simeq 0.066$. This region is populated predominantly by D/T-type asteroids, but it is outside their core distribution. Smaller numbers of dark X/P-, and Z-types objects also occur at comparable infrared-to-visible albedo ratios and geometric visible albedo values. The combined albedo and ratio comparison therefore does not uniquely associate either Phobos unit with a single asteroid class. Nevertheless, the Red Unit measurements mostly favour a similarity with the D/T-type population (and Z-types), while their overlap with the other taxonomic populations is comparatively sparse.

\subsection{Discussion}
Our analysis shows that the infrared-to-visible albedo ratio, ($R_P$), differs systematically among the three Phobos spectral units, increasing from the Blue to the Transitional and Red units. Comparison with asteroid infrared-to-visible albedo ratios ($R_A$) demonstrates that all three units overlap the parameter space occupied by primitive asteroids, with the strongest correspondence to D/T-type asteroids. Z-types also offer a potential match, but the limited number of objects in this taxonomic class does not currently allow a robust comparison. However, none of the Phobos spectral units exhibits a unique association with a single taxonomic class, as partial overlap is also found with some dark X/P-types and less so with C-complex asteroids.

The Blue Unit infrared-to-visible albedo ratios are mostly compatible with D/T-types and Z-types and to lesser extent to the dark X/P-types. This behavior is further confirmed by the position of the Blue Unit ROIs in $R_A$--$p_V$ space (Fig.~3). Our results are not necessarily inconsistent with those of \citet{wargnier2025}: these authors found that the Phobos Blue Unit shares spectral characteristics with P-type asteroids. We find that low-albedo X-types and P-types are a possible match, but less likely than D/T-types in terms of visible-to-infrared albedo ratios. However, their and our studies use different observables, wavelength ranges, and datasets. \citet{wargnier2025} compared CRISM-derived visible-to-near-infrared spectral slopes ($\lesssim 2.5~\SI{}{\micro\meter}$) together with absolute albedo, whereas our parameter $R_P$ is a normalised broad-band albedo ratio spanning the V band ($0.50\text{--}0.60~\SI{}{\micro\meter}$) and W1 band ($3.1\text{--}3.8~\SI{}{\micro\meter}$).

A complementary comparison is provided by \citet{takir2021}, who obtained disk-integrated spectra of Phobos and Deimos over approximately $0.7$--$4~\SI{}{\micro\meter}$. They found that the $0.7$--$2.52~\SI{}{\micro\meter}$ portions of both spectra were consistent with D-type asteroids in their low albedo, largely featureless character, and steep spectral slope. Our population-level results are broadly consistent with this interpretation. The Blue and Transitional Unit ROIs fall within the main D/T-type $R_A$ distribution, whereas the Red Unit ROIs occupy its higher-$R_A$ tail.
The comparison is nevertheless not direct: \citet{takir2021} analysed disk-integrated spectral shape, whereas our analysis compares spatially resolved Phobos units using a single broad-band ratio between the visible and W1-like intervals.

Under a captured-asteroid scenario, their visible-to-infrared spectral behaviour is most closely represented at the population level by the D/T-type distribution: the Blue and Transitional Unit ratios occur within its well-populated regions, while the Red Unit ratios occupy its high-$R_A$ tail. Additional overlap occurs with the dark X/P-type population and with smaller numbers of Z-type and C-complex asteroids. Detailed spectroscopy is nevertheless required to determine whether objects with similar integrated ratios also share comparable continuum shapes, spectral curvature, or absorption features. The stronger correspondence with the D/T-type population therefore does not uniquely support a captured-asteroid origin. Martian material may also contribute to the observed properties of Phobos, particularly those of the Red Unit, either directly or through mixtures produced in an impact-generated formation scenario.

The relationship between spectral redness and surface maturity is also not straightforward. Previous studies \citep{rosenblatt_2011_the, schmedemann_2014_the, ballouz2019, munaretto_2025_phase} have interpreted the red regions of Phobos as ancient and strongly space-weathered terrains. However, observations of primitive asteroids show that redder spectral behaviour does not always indicate greater surface maturity. On Bennu, a B-type asteroid, redder material has been interpreted as relatively fresh exposure \citep{dellagiustina_2020_variations}, while observations of the T-type asteroid (596) Scheila indicate that impact-exposed material can also exhibit red spectral slopes \citep{fornasier_2007_visible, avdellidou_2021_characterisation, hasegawa_2022_spectral_evolution_dark, hasegawa_2022_scheila}.

Therefore, the spectral differences between the Phobos units may reflect variations in surface maturity, compositional heterogeneity, or a combination of both. In the compositional scenario, the units could expose materials derived from different depths or lithological components of a heterogeneous source body. For example, the redder units might contain a greater proportion of spectrally red primitive material associated with the outer layers of a compositionally stratified planetesimal, whereas the bluer units might expose less-red material originating from its interior. Such an interpretation would be broadly consistent with the proposed relationship between P-, T-, D-, and Z-type asteroids and some C-types, in which the redder taxonomic classes may represent the outer shells of primitive planetesimals and C-type material their interiors \citep{vernazza2021}. However, the broad-band ratios measured here cannot by themselves establish this internal-structure scenario.

Alternatively, the different spectral units may contain varying proportions of endogenous Phobos material, Martian impact-derived material, and exogenous carbonaceous material delivered by later impacts \citep{munaretto_2025_phase,beccarelli2026,beccarelli2026b}. Such compositional mixing could influence both the broad visible-to-infrared spectral behaviour and the presence or absence of weak absorption features. The companion OMEGA analysis by \citet{beccarelli2026b} provides additional spectral context for this interpretation. In particular, the Blue Unit analysed in that study exhibits a statistically significant absorption feature at 3.37~\SI{}{\micro\meter}, interpreted as being consistent with aliphatic organic material and insoluble organic matter found in CI/CM carbonaceous chondrites, whereas the corresponding Red Unit lacks this feature. The presence of this band could therefore indicate a greater contribution of exogenous carbonaceous material to the Blue Unit.

However, laboratory irradiation experiments show that energetic processing can restructure and dehydrogenate carbonaceous materials \citep{strazzulla1992} and substantially weaken the aliphatic C--H absorption near 3.4~\SI{}{\micro\meter} \citep{godard2011}. The contrast between the Blue and Red Units may therefore instead reflect different degrees of surface processing, with the Blue Unit preserving the absorption feature because it contains comparatively fresher material. The survival of weak absorption bands following the delivery of exogenous material has also been proposed as a possible explanation for the 3~\SI{}{\micro\meter} band observed on Psyche \citep{avdellidou2018}. Consequently, the observed spectral differences may reflect either intrinsically different material mixtures or different degrees of alteration of broadly similar starting materials, and these possibilities cannot presently be distinguished unambiguously.

This illustrates an important limitation of the $R_P$--$R_A$ comparison: high infrared-to-visible ratios do not uniquely indicate a specific composition. Instead, they provide a population-level measure of integrated spectral behaviour that must be interpreted together with detailed spectroscopy.

\section{Conclusions}

We compared broad-band visible-to-infrared reflectance ratios derived for the Blue, Transitional, and Red spectral units of Phobos with asteroid $p_{\mathrm{IR}}/p_V$ values compiled in MP3C.

The Phobos units show a systematic increase in infrared-to-visible reflectance from the Blue to the Red Unit. All three lie within the overall range occupied by low-albedo primitive asteroids. The Blue and Transitional Unit ratios coincide with well-populated parts of the D/T-type distribution, while the Red Unit occupies its higher-$R_A$ tail. The Blue Unit also overlaps the upper part of the dark X/P-type distribution.

The Z-type sample formally spans the ratios of all three Phobos units, but the Transitional and Red Unit values occur within its comparatively sparse, approximately flat high-$R_A$ region rather than near a pronounced Z-type peak. The Z-type correspondence therefore represents range overlap rather than an equally strong population-level match to that of the D/T-types. Comparable ratios also occur among smaller numbers of dark X/P- and C-complex asteroids.

These similarities show that Phobos has broad-band spectral behaviour compatible with primitive asteroid surfaces. However, the ratio is not uniquely diagnostic of composition and may also reflect differences in space weathering, surface maturity, or mixtures of endogenous and exogenous material. Measurements in the $0.9$--$3.6~\si{\micro\meter}$ range by the MIRS instrument onboard MMX \citep{barucci_2021_mirs, barucci2025} will be essential for distinguishing between these possibilities.

\begin{acknowledgements}
This work is based on data provided by the Minor Planet Physical Properties Catalogue (MP3C; \url{https://mp3c.oca.eu} and \url{https://mp3c.le.ac.uk}). T.~J.~Dyer acknowledges financial support from the Centre national d'\'etudes spatiales (CNES), France, within the framework of the MMX mission and Universit\'e C\^ote d'Azur (UniCA). M.~Delbo also acknowledges support from CNES. M.~Delbo is Leverhulme Visiting Professor at the University of Leicester with financial support from the Leverhulme Trust (UK). C.~Avdellidou acknowledges support from STFC (ST/Y006062/1).
\end{acknowledgements}

\bibliographystyle{aa}
\bibliography{references.bib}

\FloatBarrier
\clearpage

\end{document}